\documentstyle[aps,prb,twocolumn,epsf,rotate]{revtex}

\begin{document}

\twocolumn[\hsize\textwidth\columnwidth\hsize\csname
@twocolumnfalse\endcsname

\title{Interlayer Coupling in Ferromagnetic Semiconductor Superlattices}

\draft

\author{T. Jungwirth$^{1,2}$, W.A. Atkinson$^{1}$, B.H. Lee$^{1}$, 
and A.H. MacDonald$^{1}$}
\address{$^{1}$Department of Physics,
Indiana University, Bloomington, Indiana 47405}
\address{$^{2}$Institute of Physics ASCR,
Cukrovarnick\'a 10, 162 00 Praha 6, Czech Republic}
\date{\today}
\maketitle

\begin{abstract}
We develop a mean-field theory of carrier-induced ferromagnetism 
in diluted magnetic semiconductors.
Our approach represents an improvement over  standard RKKY model
allowing spatial inhomogeneity of the system,
free-carrier spin
polarization, finite temperature, and free-carrier
exchange and correlation
to be accounted for self-consistently. As an example,
we calculate the electronic structure of a Mn$_x$Ga$_{1-x}$As/GaAs
superlattice with alternating ferromagnetic and
paramagnetic layers and demonstrate the possibility of
semiconductor magnetoresistance systems with
designed properties.

\end{abstract}



\vskip2pc]

Semiconductors and ferromagnetic materials play a
largely complementary role in current information processing
and storage technologies.
The possibility of exploiting synergies between
their properties
has led to interest\cite{physicstoday} in fabricating hybrid systems.
Among these, the most widely studied are ones in which
ferromagnetic metals are patterned on the surface of a
semiconductor.\cite{metalsemi}  An attractive alternative
has been presented by recent
advances\cite{tohoku,tokyo,leuven,story,haury,theory} in
the fabrication and control of diluted magnetic semiconductors  
which exhibit free-carrier induced ferromagnetism. 
In semiconductors, low intrinsic carrier densities allow
electrical transport properties to be grossly altered by
changing doping profiles or gate voltages.
In ferromagnets, it is long range order which allows weak
magnetic fields to reorient large moments and produce
significant changes in transport and other properties.  Systems
with magnetically
ordered free carriers in semiconductors open up
a plethora of intriguing possibilities for engineered electronic
properties.  Since these will generally involve spatial
patterning of dopant and magnetic ion densities, their
description requires a theory of free-carrier induced
ferromagnetism in semiconductors appropriate to inhomogeneous
systems.  In this Letter we present such a theory and apply it to
the case of layered structures which mimic metallic 
giant-magnetoresistance multilayers.\cite{magslreview}
We conclude that strong magnetoresistance effects associated 
with half-metallic (fully polarized free-carrier band) 
states can occur in such systems.

The systems in which we are interested are described by
Hamiltonians of the form
\begin{equation}
{\cal H} = {\cal H}_m + {\cal H}_f - J_{pd} \sum_{i,I} {\vec S_I}
\cdot {\vec s_i} \delta({\vec r}_i - {\vec R}_I).
\label{coupling}
\end{equation}
Here ${\cal H}_m$ is the Hamiltonian of localized spins $\vec S_I$,
located at positions ${\vec R}_I$ and
interacting with an external magnetic field and, in general,
also with each other.  ${\cal H}_f$ is the Hamiltonian of a
semiconductor free-carrier system described using an envelope
function\cite{semictext} language and including
interactions of carriers with a random disorder potential and with
other carriers.  We assume here that the semiconductor of interest
has a single parabolic band; the formalism is readily
generalized to multi-band situations which will often be of
practical interest.  The last term on the right-hand side of 
Eq.~(\ref{coupling}) represents
the exchange interaction\cite{dms} between
$\vec{S}_I$ and free-carrier spins $\vec{s}_i$ which is responsible 
for the novel physics.  Our theory is simplified by
the neglect\cite{ueda} of dynamic correlations between a localized spin and
both other localized spins and the free-carriers.
We use a grand-canonical ensemble where the full system is
in contact with a heat bath and the free carriers are in
contact with a particle reservoir.  Since we will ultimately use the
local-spin-density approximation\cite{dfreviews} (LSDA)
to account for
correlations in the inhomogeneous free-carrier system, it is
convenient to formulate our mean-field-theory in terms of the
Gibbs thermodynamic variational principle
which asserts that
\begin{equation}
\Omega \le - k_B T \ln \left[
\sum_{\alpha} \exp( - \langle \alpha | {\cal H} | \alpha
\rangle/k_B T) \right].
\label{variational}
\end{equation}
Here $|\alpha\rangle$ ranges over all possible direct products
of localized spin moment $m_I$ configurations and many-fermion free
carrier states obtained by diagonalizing ${\cal H}$ in
subspaces with fixed localized spin configuration.
\begin{equation}
\langle \alpha | {\cal H} | \alpha \rangle =
- \sum_I (g \mu_B B) m_I + E_{\kappa}[m_I]\; , 
\label{engexp}
\end{equation}
where $E_{\kappa}[m_I]$ is a many-particle eigenvalue for free-carriers
whose Hamiltonian includes the spin-dependent external potential term,
$ - \sum_i s_{z,i} [ g^* \mu_B B + J_{pd} \sum_{I} m_I
\delta (\vec r_i - \vec R_I)]$ with $g$ and $g^*$ denoting localized-spin
and free-carrier g-factors respectively.

The sum over the many-fermion
free-carrier states can be performed formally by appealing to
spin-density-functional theory\cite{dfreviews} which states that the
free-carrier grand potential can be expressed in terms of the
ground state number and spin densities.  Assuming that the total spin
quantum number of each  localized moment is $S$, it follows that 
\begin{equation}
\Omega = \sum_I \Omega^{(0)}(b_I) +
\Omega_f[n_{\uparrow},n_{\downarrow}]
- \frac{g^* \mu_B B}{2} \int d \vec r m(\vec r)
\label{omega}
\end{equation}
where $\Omega_f[n_{\uparrow},n_{\downarrow}]$ is the internal free-carrier
contribution,  
$b_I = [g \mu_B B + J_{pd} m(\vec R_I)/2]/k_B T $, and   
\begin{equation}
\Omega^{(0)}(b_I) = -k_B T \ln \big[ \sum_{i=-S}^{S} \exp (b_I i) \big]
\label{omega0}
\end{equation}
is the thermodynamic potential of an isolated
localized spin. In Eq.~(\ref{omega}), 
$m(\vec{r})=n_{\uparrow}(\vec{r})-n_{\downarrow}(\vec{r})$ and the
mean-field densities, $n_{\uparrow}(\vec{r})$ and
$n_{\downarrow}(\vec{r})$, of the itinerant spins
are to be adjusted so that the functional
derivatives of $\Omega$ with respect to both $n_{\uparrow}(\vec r)$
and $n_{\downarrow}(\vec r)$ are identically equal to the chemical
potential.

In spin-density functional theory practical
self-consistent field calculations
for inhomogeneous interacting fermions are made possible by
the Kohn-Sham separation of $\Omega_f$ into single-particle,
electrostatic and exchange-correlation pieces.  The development
here is standard apart from the introduction of an additional
spin-dependent effective potential because of the dependence of
$b_I$ on $n_{\uparrow}(\vec R_I)$ and $n_{\downarrow}(\vec R_I)$.
We find that the equilibrium free-carrier spin densities and the mean
value of the localized spin quantum numbers on each site can be
determined by solving the following equations self-consistently: 
i) a free-carrier single-particle Schr\"{o}dinger equation with
a spin-dependent potential,
\begin{equation}
\left[ - \frac{ \nabla^2}{2 m^*} + V_{\sigma}(\vec r)
\right] \psi_{k,\sigma}(\vec r)
 = \varepsilon_{k,\sigma} \psi_{k,\sigma}(\vec r)\; ,
\label{schrodinger}
\end{equation}
where
$$
V_{\sigma}(\vec r)=v_{es}(\vec r) + v_{xc,\sigma}(\vec r)
-\frac{\sigma}{2} \left(g^* \mu_B B + h_{pd}(\vec r)\right)\; ;
$$
ii) the Poisson equation for the electrostatic potential
\begin{equation}
v_{es}(\vec r) = \frac{e^2}{\epsilon} \int d \vec r' \frac{n(\vec r')}
{|\vec r - \vec r'|} + v_{ext}(\vec r)\; ,
\label{poisson}
\end{equation}
where 
$$
n(\vec r)=n_{\uparrow}(\vec r)+n_{\downarrow}(\vec r)\; , \;\;
n_{\sigma}(\vec r)= \sum_k f(\epsilon_{k,\sigma}) |\psi_{k,\sigma}(\vec
r)|^2\; ,
$$
$f(\epsilon_{k,\sigma})$ is the Fermi distribution function,
and $v_{ext}(\vec r)$ is the envelope function external
potential\cite{semictext} including most importantly
band edge and ionized impurity contributions;
iii) the LSDA equation for the 
spin-dependent exchange-correlation potential
\begin{equation}
v_{xc,\sigma}(\vec r) = \frac{d [n
\epsilon_{xc}(n_{\uparrow},n_{\downarrow})]}{d n_{\sigma}}
\big|_{n_{\sigma}= n_{\sigma}(\vec r)}\; ,
\label{xcpot}
\end{equation}
where $\epsilon_{xc}(n_{\uparrow},n_{\downarrow})$
is the exchange and correlation energy per particle of a spatially
uniform free carrier system;\cite{vosko}
iv) the mean-field equation for the exchange-coupling effective 
Zeeman field
\begin{equation}
h_{pd}(\vec r) = J_{pd} \sum_I \delta (\vec r - \vec R_I) \langle m_I
\rangle\; ,
\label{exchangefield}
\end{equation}
where the mean-field localized-spin moment
$\langle m_I \rangle = S B_S(b_I S)$ and 
$B_S(x)$ is the Brillouin function.\cite{aharoni}

\begin{figure}[b]
\epsfxsize=3.5in
\centerline{\epsffile{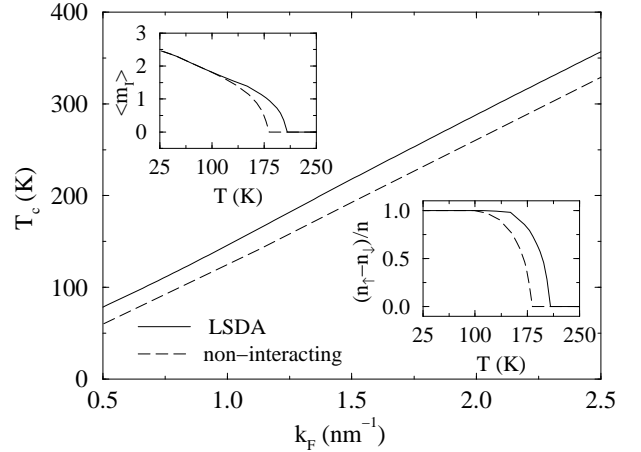}}

\vspace*{.0cm}

\caption{Curie-Weiss temperature, 
as a function of free-carrier Fermi wavevector,
calculated including (solid line) and neglecting (dashed line) 
the exchange-correlation potential.
The localized spin moment (upper inset) and the 
free-carrier relative spin polarization (lower inset) 
are plotted as a function of temperature for $k_F = 1.4$~nm$^{-1}$.}
\label{tcr}
\end{figure}

For homogeneous systems with randomly
distributed localized spins these equations can be solved  
analytically and represent an improvement over the mean-field
theory RKKY interaction description of free-carrier
induced ferromagnetism.  Our approach allows finite
temperature, free-carrier exchange and
correlation, and free-carrier spin-polarization effects to be
conveniently accounted for.  The Curie-Weiss temperature, obtained from 
Eqs.~(\ref{schrodinger})-(\ref{exchangefield}),
is given by
\begin{equation}
k_B T_c = \frac{c S (S+1)}{3} \frac{J_{pd}^2}{(g^* \mu_B)^2}
\chi_f(n,T)\; ,
\label{tc}
\end{equation}
where $c$ is the magnetic impurity density, and  
$\chi_f(n,T)$ is the temperature and 
density dependent free-carrier magnetic susceptibility.
In Fig.~\ref{tcr} we have
plotted the ferromagnetic transition temperature predicted by
this expression for $p$-type Mn$_x$Ga$_{1-x}$As
with $S=5/2$, $J_{pd}=0.15$ eV~nm$^3$,
$c=10^{21}$ cm$^{-3}$ and hole mass $m^*=0.5m_e$,
as a function of free-carrier Fermi wavevector $k_F$. 

When exchange and correlation effects are neglected
and $\chi_f(n,T)$ is replaced by its zero-temperature value,
$T_c$ is  proportional to $k_F$ and
agrees with the RKKY theory expression.\cite{theory,tohoku,ueda}
Free-carrier exchange and correlation 
enhances $T_c$ by $\approx 30~K$ in the 
range of free-carrier densities studied.  The theory appears to
be reasonably accurate when applied to the experimental 
systems we are interested in, those which have the largest
critical temperatures\cite{tohoku} 
($T_c \approx 110$~K), providing confidence in its application 
to inhomogeneous systems.  It is not adequate at small 
Mn fraction ($ x < 0.02$) 
where ferromagnetism is not observed, presumably because 
the free-carrier density is below its Mott limit.  Experimentally,
$T_c$ also decreases for $x > 0.07$, possibly because of 
spin-fluctuation effects neglected in the mean-field theory\cite{ueda}.
The insets in Fig.~\ref{tcr} show that in the density-range of interest,
the magnetization of the localized spins saturates more slowly than 
that of the free carriers as the temperature falls below the critical
temperature.
\begin{figure}[b]
\epsfxsize=3.5in
\centerline{\epsffile{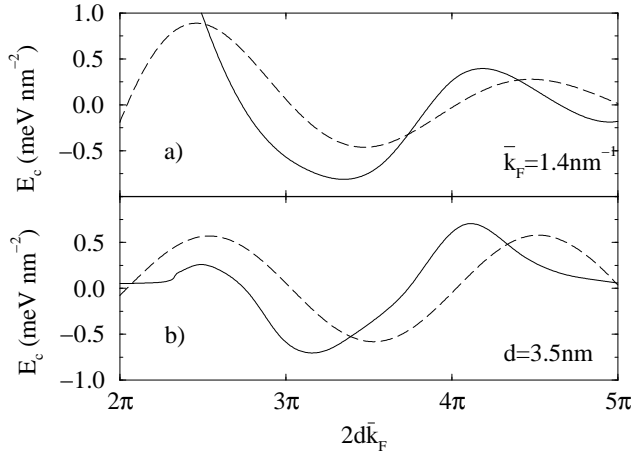}}

\vspace*{.0cm}

\caption{Interlayer exchange coupling (solid lines)
as a function of the Mn-undoped GaAs spacer thickness
(a) and as a function of the average density of free-carriers (b). 
The dashed lines is the RKKY interaction coupling energy 
calculated with all magnetic impurities confined to planes 
separated by the superlattice period. 
Results are plotted as a function of dimensionless parameter 
$2d\overline{k}_F$.
}
\label{ec}
\end{figure}

Once the band edge and ionized and magnetic impurity
profiles have been specified,
Eqs.~(\ref{schrodinger})-(\ref{exchangefield}) 
can be used to solve for the sytem's equilibrium properties.
As an illustration of our approach, we consider a  Mn$_x$Ga$_{1-x}$As/GaAs
superlattice  with magnetic impurities in alternate layers. 
We look for two different self-consistent solutions of 
Eqs.~(\ref{schrodinger})-(\ref{exchangefield}), a ferromagnetic (F) one with 
parallel ordered moments
in all Mn-doped regions, and a solution with 
an antiferromagnetic (AF) alignment of adjacent magnetic layers. The interlayer
exchange coupling $E_c$, is defined as the difference in energy between
AF and F-states per area 
per Mn$_x$Ga$_{1-x}$As layer and is expected\cite{rkky} to
oscillate with GaAs spacer width.  In 
Fig.~\ref{ec}  we present numerical results for $E_c$ as a function
of a dimensionless parameter $2d\overline{k}_F$ where $\overline{k}_F$
is the Fermi wavevector corresponding to the average 3D density of free carriers
in the superlattice with a period $d$. The system we consider has
2~nm thick Mn$_x$Ga$_{1-x}$As layers with $c=10^{21}$ cm$^{-3}$
and  homogeneously distributed ionized impurities
that neutralize free-carrier charge. Oscillations in the self-consistent
(solid lines) $E_c$, are 
qualitatively consistent with simple RKKY model estimates;\cite{rkky} 
differences exist primarily because of the proximity induced spin-polarization
in the nominally paramagnetic GaAs layers mentioned below.
The amplitude of oscillations in $E_c$
is $\sim 10\times$ smaller than in metallic systems\cite{ecmetal}
measured in absolute units and $\sim 10\times$ larger if energy is
measured relative to the Fermi energy of free carriers.
In order to achieve substantial exchange coupling in experimental
systems it will be important to limit disorder scattering in the 
GaAs layers.  
\begin{figure}[b]
\epsfxsize=3.5in
\centerline{\epsffile{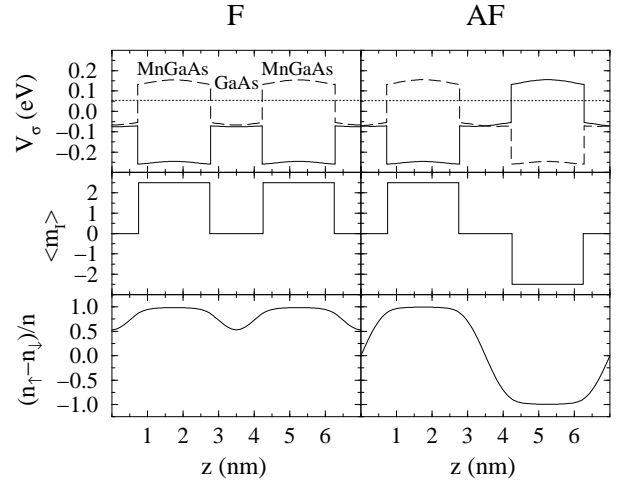}}

\vspace*{-.0cm}

\caption{\protect
Self-consistent results plotted within the unit cell of
the Mn$_x$Ga$_{1-x}$As/GaAs
superlattice. From top to bottom: i) effective potentials from 
Eq.~\ref{schrodinger} for up (solid lines) and down (dashed lines)
spin free-carriers (the dotted line here indicates the chemical potential); 
ii) mean-field localized spin moment; iii) 
free-carrier relative spin-polarization. 
}
\label{vmixi}
\end{figure}

\begin{figure}[b]
\epsfxsize=3.5in
\centerline{\epsffile{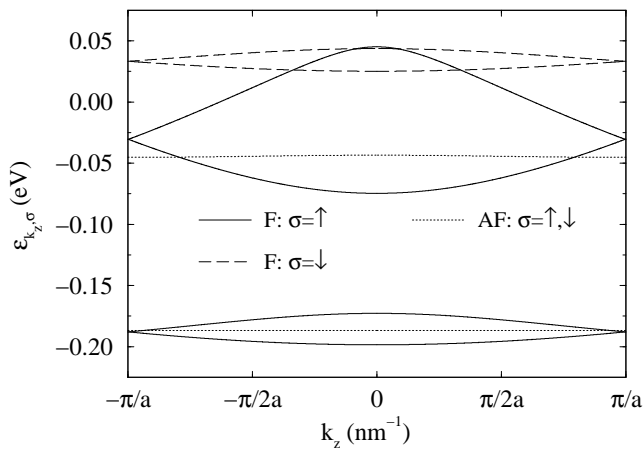}}

\vspace*{.0cm}

\caption{Partially occupied energy minibands in F-state for spin-up (solid line)
and spin-down (dashed line).  In the AF-state both spins (dotted line)
have the same minibands.  The chemical potential is 0.053~eV and 
$a$=7~nm is the unit cell length.}
\label{e}
\end{figure}

Our mean-field calculation also yields information on the 
localized and free-carrier spin magnetization densities and on the spin-split
bands of the free-carrier system. This information
provides a starting point
for building a theory of electronic transport. Numerical results for the
above Mn$_x$Ga$_{1-x}$As superlattice with $d=3.5$~nm and 
$\overline{k}_F=1.4$~nm$^{-1}$ (corresponding to average density 
10$^{20}$~cm$^{-3}$) are summarized in Figs.~\ref{vmixi} and \ref{e}.
In the top panels of Fig.~\ref{vmixi}, the chemical potential and 
the effective potentials 
for spin-up and spin-down free carriers (see Eq.~(\ref{schrodinger}))
are plotted as a function of $z$ over a AF configuration unit cell.
All energies here and below are measured from the
spatial average of the electrostatic potential $v_{es}(z)$.
The potentials $V_{\sigma}$ have similar shapes in F and AF cases,
except for the reversed order for up and down spins in the right
Mn-doped layer.  Note that in this example, confinement of 
carriers in the magnetic layers is due entirely to the exchange potential
produced by magnetic order.  The  localized  (middle panels)
and itinerant (lower panels) spin systems reach 100~$\%$
polarization in the Mn-doped layers at the 
temperature ($T=0.1T_c$) and carrier density for which these
calculations were performed.  
The itinerant system spin-polarization is large 
in the layers free of magnetic impurities, especially 
so in the F-state case. 

Fig.~\ref{e} shows occupied minibands 
in the superlattice Brillouin zone.
In the F-state spin-up and spin-down minibands are split by about 0.25~eV.
There is no spin-splitting in the AF-state since the  effective
potentials $V_{\uparrow}$ and $V_{\downarrow}$ differ by a rigid 
shift in the $\hat{z}$ direction. The miniband dispersion is
much weaker in the AF case because the barriers separating
two adjacent minima in the effective potential are twice as thick and 
high as in the F case.
Since the conductance is approximately proportional to the square 
of the largest miniband width in either coherent or incoherent 
transport limits, the minibands can be used to estimate 
the size of the current-perpendicular-to-plane (CPP) magnetoresistance  
effect.  For the case illustrated, the AF state CPP conductance 
will be three orders of magnitude smaller than the F state CPP
conductance. The large difference 
is expected since the
bulk Mn$_x$Ga$_{1-x}$As bands are half-metallic
for these parameters.  In general we predict strong CPP magnetoresistance
in Mn$_x$Ga$_{1-x}$As/GaAs multilayer systems if the AF state can be realized.

The GaAs/Mn$_x$Ga$_{1-x}$As superlattice is but one example of a
semiconductor nanostructure to which our mean field theory 
can be applied.  Tremendous flexibility is possible even within systems
containing only GaAs based materials.
These can act like a normal (N)
conductor, like an itinerant ferromagnet (F) when a fraction of 
the Ga atoms is replaced by Mn, and like
an insulator (I) when a large enough fraction of Ga in the crystal is 
replaced by Al.  In GaAs-based N/F/I heterostructures all components
are lattice matched and have relatively simple band structures.  
Geometries of interest include F-N junctions, N-F-N spin-filters,
F-N-F spin-valves, and F-I-F magnetic tunnel junctions. Since the materials
are semiconductors rather than metals, external bias voltages can have
a strong influence on both charge and magnetization density profiles.
Self-consistency is hence essential in modeling these systems.

The authors acknowledge helpful interactions
with D.D. Awaschalom, J.K. Furdyna. and E. Miranda. 
This work was supported by the National Science Foundation under
grants DMR-9623511, DMR-9714055 and INT-9602140,
by the Ministry of Education of the Czech Republic under
grant ME-104 and by the Grant Agency of the Czech Republic
under grant 202/98/0085.

\end{document}